\documentclass[universe,article,pdftex,moreauthors]{Definitions/mdpi} 

\firstpage{1} 
\pubvolume{1}
\issuenum{1}
\articlenumber{0}
\pubyear{2022}
\copyrightyear{2022}
\datereceived{} 
\dateaccepted{} 
\datepublished{} 

\renewcommand{\linenumbers}{}

\hreflink{https://doi.org/} 

\Title{Construction and evolution of equilibrium configurations of the Schr\"odinger-Poisson system in the Madelung frame}

\TitleCitation{SP system in Madelung frame}

\Author{Iv\'an Alvarez-R\'ios$^{\dagger}$\orcidA{} and Francisco S, Guzm\'an$^{\ddagger}$\orcidB{}}

\AuthorNames{Firstname Lastname, Firstname Lastname and Firstname Lastname}

\AuthorCitation{Alvarez-R\'ios, I.; Guzm\'an, F. S.}

\address{%
$~~~$\quad Instituto de F\'{\i}sica y Matem\'{a}ticas, Universidad
              Michoacana de San Nicol\'as de Hidalgo. Edificio C-3, Cd.
              Universitaria, 58040 Morelia, Michoac\'{a}n,
              M\'{e}xico
}

\firstnote{ivan.alvarez@umich.mx} 
\secondnote{francisco.s.guzman@umich.mx}

\abstract{We present the construction of ground state equilibrium configurations of the Schr\"odinger-Poisson (SP) system in the Madelung frame and evolve such configuration using finite volume methods. We compare the behavior of these configurations when evolved within the SP and Madelung frames, in terms of conservation of mass and energy. We also discuss the issues of the equations in the Madelung frame and others inherent to the numerical methods used to solve them.}

\keyword{dark matter -- Bose condensates} 

\begin{document}


\section{Introduction}

The Schr\"odinger-Poisson (SP) system of equations has recently called attention because these equations rule the dynamics of the Fuzzy Dark Matter model (FDM), which proposes that dark matter is an ultralight spinless boson (e.g. \citep{Matos-Urena:2000,Sahni:2000,Hu:2000,Hui:2016}). In this model, Schr\"odinger equation plays the role of the Gross-Pitaevskii equation for a Bose gas whereas Poisson equation provides the potential trap that contains the bosons, whose source is the bosonic gas density itself. The solution of the SP system of equations is the essence of the analyses of this dark matter candidate.

Among the most interesting discoveries within this model is that structures resulting from the evolution of dark matter fluctuations, accommodate in density configurations with a core and tail (e.g. \cite{Schive:2014dra,Mocz:2017wlg}). The core happens to have the profile of ground state equilibrium configurations of the SP system \cite{Ruffini:1969,GuzmanUrena2004}, which have been found, in isolated scenarios, to be late-time attractor solutions \cite{GuzmanUrena2006,BernalGuzman2006b}.

Now, on the numerical aspect of the simulations based on the numerical solution of the SP system under various astrophysical scenarios, it happens that different numerical methods and approaches are used, see e. g. \cite{JiajunZhang,Niemeyer_2020} for reviews on the subject. The various analyses use two main frames for the solution of the SP system, some of them consist on the direct solution of the SP equations, whereas others solve the hydrodynamical, Madelung version of the equations \cite{Madelung,ChavanisMadelung}. 

Examples solving the SP system include \cite{2009ApJ...697..850W}, where the first structure formation with ultralight dark matter was experimented with, \cite{Schive:2014dra} where the structure formation within the FDM model using high resolution simulations receives an important boost, in \cite{Guzman2019} the oscillation mode spectrum of cores is studied, in
\cite{BernalGuzman2006a} the solitonic behavior of cores is revised, the attractor nature of cores is first presented in \cite{BernalGuzman2006b} for the collapse of non-spherical fluctuations, in
\cite{PhysRevLett.123.141301} the star forming structures in FDM Filaments is presented, in 
\cite{Gotinga2022} zoom in simulation of a galactic halo formation solves the SP equations together with SPH methods, 
simulations of core mergers are developed in \cite{Schwabe:2016} and in \cite{PhysRevD.97.063507} tidal disruption of FDM subhalo cores is studied.

On the other hand equations in the madelung frame have been used to
study various other problems, for example, the core-halo mass relations and the possible formation of supermassive black holes \cite{ShapiroCreTail}, in \cite{Schobesberger_2021} the existence of vorticity is analyzed in core structures, in 
\cite{Li_2021} local properties of cores is studied, including random motion and collisions, in  
\cite{Shapiro2021} the Jeans mass shrinking is studied within the scalar field dark matter model, the merger of cores was seen in \cite{Maleki_2020}, in \cite{PhysRevE.91.053304} the equations are solved using the SPH method and in \cite{Mocz:2018} the Schr\"odinger-Poisson Vlasov-Poisson correspondence is proposed.

Interesting issues emerge in the Madelung frame though, first of all is that Schr\"odinger equation is cast in an analog fashion to Euler equations, and therefore the methods developed for the evolution of fluids can be applied. On the other hand, an important drawback of the Madelung frame is that unlike Euler evolution equations of a compressible fluid, there is not an equation for the balance of energy, and there is no clear Equation of State (EoS) for this quantum fluid analog. Various proposals  starting from the Boltzmann equation indicate an EoS for example in \cite{Shapiro2021} and  \cite{Mocz:2018} where the SP system is identified with the Poisson-Vlasov system. In our approach below, no EoS or microscopic approach is considered.

Amid the important advances in FDM simulations, the comparison of numerical solutions in the SP and Madelung frames are needed in order to learn about the pros and cons of each approach, and some studies have started in the structure formation scenario  \cite{PhysRevD.99.063509}. In this paper we go slightly back and practice a comparison in the simplest of scenarios where the two frames can be compared, namely, the construction and evolution of ground state equilibrium configurations, which, as said before, plays an essential role as galactic cores in the FDM astrophysics. We construct the ground state equilibrium configurations directly on the Madelung frame and evolve them with a code that implements methods used in hydrodynamics in order to compare the dynamics in the two frames.

The paper is organized as follows. In Section \ref{sec:eqs} we rewrite the SP equations in the Madelung frame and construct the ground state equilibrium solutions. In Section \ref{sec:evolution} we draw a numerical method suitable for the evolution of these configurations and compare the dynamics of equilibrium configurations in the two frames. Finally in Section \ref{sec:conclusions} we draw some conclusions.

\section{Madelung transform}
\label{sec:eqs}

We start by writing the SP equations in the FDM regime:

\begin{eqnarray}
i \frac{\partial \Psi}{\partial t} &=& -\frac{1}{2}\nabla^2 \Psi + V\Psi ,\label{eq:SchroNoUnits}\\
\nabla^2 V &=& 4\pi |\Psi|^2, \label{eq:PNoUnits}
\end{eqnarray}

\noindent where $\Psi$ is the wave function or parameter order of the boson gas and $V$ is the potential trap, which is the gravitational potential sourced by the gas mass itself. These equations are already scaled so that constants are absorbed.

The Madelung transformation defines the wave function in terms of new variables $\rho,v, S$, $ \Psi = \sqrt {\rho} e^{iS} $, and transforms Schr\"odinger equation into \cite{Madelung}:

\begin{eqnarray}
\partial_t\rho &+& \nabla\cdot\left(\rho\vec{v}\right)=0,\nonumber\\
\partial_t S &+& \dfrac{1}{2} |\nabla S|^2 + V + Q = 0,\nonumber
\end{eqnarray}

\noindent equations for $\rho$ and the phase $S$. By further defining $\vec{v} = \nabla S$, the SP system is finally transformed into the constrained evolution system

\begin{eqnarray}
\partial_t\rho &+& \nabla\cdot\left(\rho\vec{v}\right)=0,\label{eq: mass conservation}\\
\partial_t\vec{v} &+& \vec{v}\cdot\nabla\vec{v} = -\nabla\left(Q+V\right),\label{eq: momentum conservation}\\
\nabla^2 V & = & 4\pi \rho,\label{eq: Poisson Madelung}
\end{eqnarray}

\noindent where

\begin{equation}
Q = -\dfrac{1}{2}\dfrac{\nabla^2\sqrt{\rho}}{\sqrt{\rho}}
\label{eq:qpotential}
\end{equation}

\noindent is known as the quantum potential. In this frame $\rho$ is interpreted as the density of a fluid and $\vec{v}$ as its velocity field. Equations (\ref{eq: mass conservation}) and (\ref{eq: momentum conservation}) correspond to the transformed Schr\"odinger equation, which looks now as a set of two flux balance laws corresponding to the mass conservation and a momentum density balance, analog to Euler equation for a fluid, except by the quantum potential. Notice that unlike the fluid dynamics equations, there is no equation for the balance of an energy neither an equation of state for the fluid, although in some regimes it is possible to identify the fluid with a polytrope with adiabatic index $\Gamma=2$ \cite{Chavanis_2019}.

\subsection{Diagnostics}

The quantities to monitor in both frames are the mass of the system $M$, and its total energy $E = K + W$, where $K$ and $W$ are the kinetic and potential energies defined in each frame by

\begin{equation}
M = \int |\Psi|^2 d^3x = \int \rho d^3x ,\label{eq:mass}
\end{equation}

\begin{eqnarray}
K &=& -\frac{1}{2}\int\Psi^*\nabla^2\Psi d^3x \nonumber\\
&=& \frac{1}{2}\int\rho|\vec{v}|^2 d^3x + \frac{1}{2}\int|\nabla\sqrt{\rho}|^2d^3x,
\end{eqnarray}

\begin{equation}
W = \frac{1}{2}\int\Psi^*V\Psi d^3x = \frac{1}{2}\int\rho V d^3x,
\end{equation}

\noindent that hold when integrated in the spatial domain.

\subsection{Ground state equilibrium configurations in the SP frame}

We briefly summarize the well known construction of equilibrium configurations in the SP frame, extensively described in e. g. \cite{Ruffini:1969,GuzmanUrena2004}. The first assumption is spherical symmetry, and thus spherical coordinates $(r,\theta,\phi,t)$ are appropriate, so that the wave function depends on the radial coordinate and time $\Psi=\Psi(r,t)$. The second assumption is that the wave function depends harmonically of time, that is $\Psi(r,t)=e^{i\omega t}\psi(r)$, which ensures $|\Psi|^2$ is time independent and consequently the gravitational potential as well. Equations (\ref{eq:SchroNoUnits})-(\ref{eq:PNoUnits}) are reduced to

\begin{eqnarray}
\frac{d^2 \psi}{dr^2} &+& \frac{2}{r}\frac{d\psi}{dr}=2(V+\omega)\psi,\label{eq:SPSchro}\\
\frac{d^2 V}{dr^2} &+& \frac{2}{r}\frac{dV}{dr} = 4\pi |\psi|^2,\label{eq:SPPoisson}
\end{eqnarray}

\noindent which define an eigenvalue problem for $\psi(r)$ provided suitable boundary conditions for the wave function $\psi(0)=\psi_c$ and $d\psi/dr(0)=0$ at the origin, and isolation, which means $\psi(r \rightarrow \infty)\rightarrow 0$ and $d\psi/dr(r \rightarrow \infty)\rightarrow 0$. For the gravitational potential the conditions at the origin are $V(0)=V_c$ , $dV/dr(0)=0$ and at infinity a monopolar condition is used $V(r \rightarrow \infty)=-M/r$ where $M=\int|\psi|^2 r^2 dr$. Fulfillment of these conditions imply that for each central value of the wave function $\psi_c$ there is a unique eigenfrequency $\omega$ that satisfies the equations. In this sense, one can construct a one parameter family of solutions labeled by the central value $\psi_c$.

Finally, unlike excited state solutions, ground state configurations are characterized by the condition that $\psi(r)$ has no zeroes within the domain of integration, otherwise solutions that satisfy the boundary conditions and has zeroes, are known as excited states (see e.g. \cite{Ruffini:1969,GuzmanUrena2004,GuzmanUrena2006}).

This eigenvalue problem is solved on a discrete domain $r\in[0,r_{max}]$ using the shooting method as described in \cite{GuzmanUrena2004}, with a tolerance on the fulfillment of boundary conditions at large radius. The solution for $\psi_c=1$ in Fig. \ref{fig:equilibrium}.

\subsection{Ground state equilibrium configuration in the Madelung frame}

These are spherically symmetric stationary solutions of the system (\ref{eq: mass conservation})-(\ref{eq: Poisson Madelung}). For the construction of a stationary configuration we also assume spherical symmetry and use spherical coordinates. The flow is assumed to be stationary, which drops the time derivatives in equations (\ref{eq: mass conservation})-(\ref{eq: momentum conservation}). The mass conservation reduces to an identity whereas the momentum density balance becomes the equation

\begin{equation}
\dfrac{d}{dr}\left(Q(r) + V(r)\right) = 0,
\end{equation}

\noindent which can be integrated immediately to obtain a constraint between the gravitational and quantum potentials $Q(r) + V(r) = V_0$, where $V_0$ is a constant of integration. Considering the quantum potential is given by (\ref{eq:qpotential}), this equation is a second-order differential equation for the density $ \rho $:

\begin{equation}
\rho''(r) = 4(V(r)-V_0)\rho(r) - \dfrac{2\rho'(r)}{r}+\dfrac{\rho'(r)^2}{2\rho(r)}.
\label{eq: rho biprime}
\end{equation}

\noindent Notice that substitution of $\rho = \sqrt{\psi}$ in this equation gives the stationary Schr\"odinger equation (\ref{eq:SPSchro}) with $\omega = - V_0$. In order to solve this equations we write it down as a first-order system by defining $u (r) = \rho'(r)$. In this way  Eq.  (\ref{eq: rho biprime}) becomes

\begin{eqnarray}
u'(r) &=& 4(V(r)-V_0)\rho(r) - \dfrac{2 u(r)}{r}+\dfrac{u(r)^2}{2\rho(r)},\label{eq: u prime}\\
\rho'(r) &=& u(r).\label{eq:rhoprima}
\end{eqnarray}

\noindent Now, Poisson equation (\ref{eq: Poisson Madelung}) is rewritten also as a first order system using the definition of $m'(r)=4\pi r^2 \rho(r)$ which leads to the equations

\begin{eqnarray}
m'(r) &=& 4\pi r^2 \rho(r), \label{eq: mass}\\
V'(r) &=& \dfrac{m(r)}{r^2}.\label{eq:potprima}
\end{eqnarray}

\noindent The system (\ref{eq: u prime})-(\ref{eq:potprima}) is the set of equations for equilibrium configurations to be solved. However notice that their form is not suitable for numerical integration at the origin. Expansion of $u'(r)$ near zero gives:

\begin{eqnarray}
\lim_{r\to 0} u'(r) & = & \lim_{r\to 0}\left(4(V(r)-V_0)\rho(r) - \dfrac{2 u(r)}{r}+\dfrac{u(r)^2}{2\rho(r)} \right)\nonumber\\
& = & 4(V_c - V_0)\rho_c - 2 \lim_{r\to 0}\dfrac{u(r)}{r} \nonumber\\
& = & 4(V_c - V_0)\rho_c - 2\lim_{r\to 0}u'(r)~~~~~\Rightarrow\nonumber\\
\lim_{r\to 0} u'(r) &=& \dfrac{4}{3}(V_c - V_0)\rho_c,\label{eq: u primeorigin}
\end{eqnarray}

\noindent is the equation to be solved at the origin instead of (\ref{eq: u prime}).

The boundary conditions imposed to these equations are $m(0)=0$, $\rho(0)=\rho_c$, $\rho'(0)=u(0)=0$ and $V(0) = V_c$, a value that can be  arbitrary, and at infinity $\lim_{r\to\infty} \rho(r) = \rho_\infty$, $\lim_{r\to\infty}u(r)=0$. With these conditions the system (\ref{eq: u prime})-(\ref{eq: u primeorigin}) becomes an eigenvalue problem for the eigenvalue $V_0$.

We numerically integrate the system and approximate the value of $V_0$ using the Shooting method, on a discrete domain $r\in[0,r_{max}]$, with the condition of minimizing the value of the function $E(V_0) = \frac{1}{2}(\rho(r_\infty;V_0)-\rho_\infty)^2 + \frac{1}{2}u(r_\infty;V_0)^2$, where $\rho_\infty$ is a finite small value that helps approximate the solution to zero at infinity with finite numerical precision.

\begin{figure}[hbtp]
\centering
\includegraphics[width=7cm]{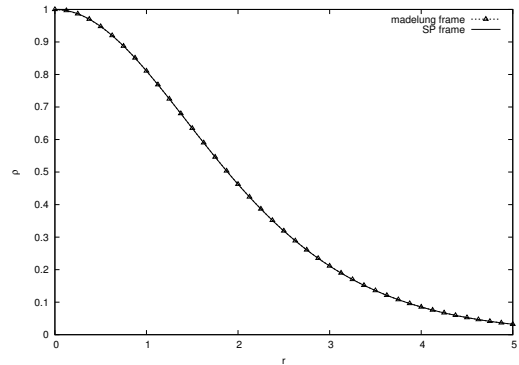}
\caption{Densities $|\psi|^2$ and $\rho$ of the ground state equilibrium solution in the SP and the Madelung frames respectively. In the SP frame equations (\ref{eq:SPSchro})-(\ref{eq:SPPoisson}) are solved for the central wave function $\psi_c=1$. In the Madelung frame equations (\ref{eq: u prime})-(\ref{eq: u primeorigin}) we use the central density $\rho_c=1$. The numerical solution is constructed on a discrete domain with resolution $\Delta r = 2.5\times 10^{-4}$.}
\label{fig:equilibrium}
\end{figure}

In Figure \ref{fig:equilibrium} we show a zoom of the density profile $\rho$ obtained for the solution with central density $\rho_c = 1$, and $\rho_\infty = 0$ within a small tolerance. Superposed we show the solution of the eigenvalue problem (\ref{eq:SPSchro})-(\ref{eq:SPPoisson}) for $\psi_c=1$. The eigenvalue of the solution in the two frames is $-V_0=\omega=-0.6922$.

\section{Evolution}
\label{sec:evolution}

A further comparison between the solutions in the SP and Madelung frames is the evolution of these configurations. In fact using one or the other frame motivates the use of different numerical methods. In the SP frame it is common to use Finite Difference methods, with a variety of time integrators, explicit or implicit, because Schr\"odinger equation is dispersive, which prevents the formation of discontinuities. In the contrary, Eq. (\ref{eq: momentum conservation}) is quasilinear, which may lead to the formation of discontinuities and shocks even for smooth initial data, and then other numerical methods are needed. 

In the SP frame it is possible to show that when solving (\ref{eq:SchroNoUnits})-(\ref{eq:PNoUnits}) the wave function $\Psi$ oscillates with a frequency that coincides with $\omega$ obtained from the solution of the eigenvalue problem of Eqs. (\ref{eq:SPSchro})-(\ref{eq:SPPoisson}). Nevertheless in the Madelung frame there is not an equivalent diagnostics.

A very important aspect is the differentiation of the wave function $\Psi$, since in general it can be considered for an equilibrium configurations as a smooth function. On the other hand, in the Madelung framework some of the variables are not even continue. We can see this from the definition of $S$, namely the argument of the function $\Psi$, which is not defined at the origin, and therefore for all cases where $S$ is not spatially constant, it will lead to initial data with discontinuous velocities that will tend to produce shock waves, which do not appear in the SP framework. 
A discontinuity of the velocity leads to a discontinuity of the density, and consequently $Q$ and $\nabla Q$ are undefined for the right hand side of Eq. (\ref{eq: momentum conservation}).
Only in cases where the function $S$ is constant can the evolution lead to a solution, even a weak solution.

We then consider that problems in which SP and Madelung frames can be compared, need to be those with a constant velocity field. In what follows we describe a comparison of the evolution of the ground state configuration in each of the frames, accompanied with a description of the respective numerical methods.

To start the evolution we interpolate the equilibrium solutions from Figure \ref{fig:equilibrium}, constructed in spherical coordinates into a 3D cubic domain described with Cartesian coordinates $D=[x_{min},x_{max}]^3$, uniformly discretized with $N$ cells along each direction that defines the numerical domain $D_d=\{ (x_i,y_j,z_k) | x_i=x_{min}+i\Delta x, y_j=x_{min}+j\Delta x, z_k=x_{min}+j\Delta x\}$, $j=0,...,N$, where $\Delta x = \frac{x_{max}-x_{min}}{N}$ is the spatial resolution along the three directions and $\Delta t=CFL \Delta x^2$ is the time resolution.
In the scenarios explored, the domain $D = [-20,20]^3$ is used with a resolution $\Delta x = 0.4$ and a courant number $CFL = 0.25$.

Once the equilibrium configurations in the SP frame is interpolated into the 3D domain described in Cartesian coordinates, we solve the system (\ref{eq:SchroNoUnits})-(\ref{eq:PNoUnits}) to evolve the configuration in the SP frame using appropriate numerical methods. Likewise, when the equilibrium configuration constructed within the Madelung frame is interpolate into the 3D domain we solve equations (\ref{eq: mass conservation})-(\ref{eq:qpotential}).

\subsection{Evolution in the SP frame}. 

The wave function evolves according to Schr\"odinger equation (\ref{eq:SchroNoUnits}), whereas Poisson equation (\ref{eq:PNoUnits}) is a constraint. For the evolution two methods are common and implemented here: the Method of Lines (MoL) and the implicit Cranck-Nicholson (CN). 

For the {\it MoL}, the semi-discrete version  of the equations uses Finite Differences with second order accurate  spatial derivatives in Schr\"odinger equation:

\begin{eqnarray}
\frac{\partial \Psi}{\partial t} &=& 
\frac{i}{2}\frac{\Psi_{i+1,j,k}-2\Psi_{i,j,k}+\Psi_{i-1,j,k}}{\Delta x^2} \nonumber\\
&+&\frac{i}{2}\frac{\Psi_{i,j+1,k}-2\Psi_{i,j,k}+\Psi_{i,j-1,k}}{\Delta y^2} \nonumber\\
&+&\frac{i}{2}\frac{\Psi_{i,j,k+1}-2\Psi_{i,j,k}+\Psi_{i,j,k-1}}{\Delta z^2} 
-i V\Psi_{i,j,k}
\end{eqnarray}

\noindent where $\Psi_{i,j,k}$ is the wave function at point $(x_i,y_j,z_k)\in D_d$ at time $t^n$. We integrate this equation from time $t^n$ to time $t^{n+1}$ using a third order Runge-Kutta explicit integrator.

The {\it Crank-Nicolson} method assumes that the evolution of the wave function from time $t^n$ to time $t^{n+1}$ is constructed as follows

\begin{eqnarray}
\Psi^{n+1}_{i,j,k} &=& \frac{e^{-{\rm i}\frac{1}{2}\hat{H}\Delta t}}{e^{{\rm i}\frac{1}{2}\hat{H}\Delta t}}\Psi^n_{i,j,k} ~~~\Rightarrow\nonumber\\
{\rm i}\frac{\Psi^{n+1}_{i,j,k}-\Psi^{n}_{i,j,k}}{\Delta t} &=& \frac{1}{2}\left[ \hat{H}\Psi^{n+1}_{i,j,k}+ \hat{H}\Psi^{n}_{i,j,k}\right]\label{eq:3p1ADI}.
\end{eqnarray}

\noindent where we add the upper label to the wave function indicating the time. We solve for $\Psi^{n+1}_{i,j,k}$ using the Alternating Direction Implicit (ADI) strategy, that splits the application of the Hamiltonian $\hat{H}=-\frac{1}{2}\left(\frac{\partial^2}{\partial x^2}+\frac{\partial^2}{\partial y^2}+\frac{\partial^2}{\partial z^2}\right) +V$, along each of the three spatial dimensions as follows:

{\small
\begin{eqnarray}
\left( 1-\frac{\rm i}{4} \Delta t \frac{\partial^2}{\partial x^2}\right) R_{i,j,k} &=& 
\left( 1+\frac{\rm i}{4} \Delta t \frac{\partial^2}{\partial x^2}\right) \Psi^{n}_{i,j,k}\nonumber\\
\left( 1-\frac{\rm i}{4} \Delta t \frac{\partial^2}{\partial y^2}\right) S_{i,j,k} &=& 
\left( 1+\frac{\rm i}{4} \Delta t \frac{\partial^2}{\partial y^2}\right) R_{i,j,k}\nonumber\\
\left( 1-\frac{\rm i}{4} \Delta t \frac{\partial^2}{\partial z^2}\right) T_{i,j,k} &=& 
\left( 1+\frac{\rm i}{4} \Delta t \frac{\partial^2}{\partial z^2}\right) S_{i,j,k}\nonumber\\
\left( 1+\frac{\rm i}{2} \Delta t ~V\right) \Psi^{n+1}_{i,j,k} &=& 
\left( 1-\frac{\rm i}{2} \Delta t ~V\right) T_{i,j,k}\label{eq:ADISchro3D}
\end{eqnarray}
}

\noindent where $R_{i,j,k}$, $S_{i,j,k}$ and $T_{i,j,k}$ are auxiliary numbers that store the values of the wave function after applying the derivative operator along each of the spatial directions. Each of the three first equations defines a tridiagonal system of equations when derivatives are discretized:

{\small
\begin{eqnarray}
&&(-\alpha)R_{i-1,j,k} + (1+2\alpha)R_{i,j,k} + (-\alpha)R_{i+1,j,k} =\nonumber\\
&&~~~~(\alpha)\Psi^{n}_{i-1,j,k}+(1-2\alpha)\Psi^{n}_{i,j,k} + (\alpha)\Psi^{n}_{i+1,j,k},\nonumber\\
&&(-\alpha)S_{i,j-1,k} + (1+2\alpha)S_{i,j,k} + (-\alpha)S_{i,j+1,k}  =\nonumber\\
&&~~~~(\alpha)R_{i,j-1,k}+(1-2\alpha)R_{i,j,k} + (\alpha)R_{i,j+1,k},\nonumber\\
&&(-\alpha)T_{i,j,k-1} + (1+2\alpha)T_{i,j,k} + (-\alpha)T_{i,j,k+1}= \nonumber\\
&&~~~~(\alpha)S_{i,j,k-1}+(1-2\alpha)S_{i,j,k} + (\alpha)S_{i,j,k+1},\nonumber\\
&&\Psi^{n+1}_{i,j,k} = \frac{1- \beta  V^{n+1/2}_{i,j,k}}{1+ \beta V^{n+1/2}_{i,j,k}}T_{i,j,k},\label{eq:ADISchro3Dtrid}
\end{eqnarray}
}

\noindent where $\alpha=\frac{1}{4}{\rm i}\frac{\Delta t}{\Delta x^2}$ and $\beta=\frac{1}{2}{\rm i}\Delta t$, that we solve using forward and backward substitution. Notice that the potential is evaluated at the intermediate time $V^{n+1/2}$, which is important because the potential is time-dependent. We calculate this potential as the average $V^{n+1/2}=\frac{1}{2}(V^{n+1}+V^n)$, where $V^{n+1}$ is calculated by solving the system (\ref{eq:ADISchro3Dtrid}) for $\Psi^{n+1}$, then solving Poisson equation for the source $|\Psi^{n+1}|^2$ to obtain $V^{n+1}$, going back the time-step, and integrating in time with the averaged potential.

In both MoL and CN methods, Poisson equation is solved using a Multigrid method that uses a 2-levels V-cycle as in \cite{GuzmanAlvarezGonzalez2021}. For the evolution of the equilibrium configuration we impose a boundary condition consistent with the isolation of the system and use a monopolar boundary condition for the gravitational potential $V_{\partial D}=-M/r_{\partial D}$, where $r_{\partial D}$ is the distance from the origin to a given point of the numerical boundary and $M$ is given by Eq. (\ref{eq:mass}). 

For the wave function we implement a sponge that absorbs the modes approaching the faces of the boundary and prevents their reflection. This sponge consists in the addition of an imaginary potential that acts as a sink as described in \cite{GuzmanUrena2004,Grupo2014}. Briefly, the potential $V$ in (\ref{eq:SchroNoUnits}) is redefined as $V\rightarrow V+V_{im}$, where $V$ is the solution of (\ref{eq:PNoUnits}) and $V_{im}$ is a spherical function with tanh-profile that goes from zero within a sphere containing the region where the interesting physics happens and minus one near the boundary of the domain, which absorbs the wave function if it approaches the boundary.

\subsection{Methods for the Madelung frame.} 

In this frame the evolution of the system for density $\rho$ and velocity field $\vec{v}$ is ruled by  equations (\ref{eq: mass conservation}) and (\ref{eq: momentum conservation}), whereas again Poisson equation (\ref{eq: Poisson Madelung}) is a constraint that has to be fulfilled at each time step during the evolution of density and velocity.

We use a Finite Volume discretization to solve the system (\ref{eq: mass conservation})-(\ref{eq: momentum conservation}) following  \cite{2009JCoPh.228.1713L}. Essential to the method is the appropriate construction of numerical fluxes and the characteristic structure of the equations, which can be cast in the form

\begin{equation}
\frac{\partial \vec{u}}{\partial t} + \nabla \cdot \vec{F} = \vec{S},\label{eq:fluxbalance}
\end{equation}

\noindent where $\vec{u}$ is a vector of conserved variables, $\vec{F}=[\vec{F}_x,\vec{F}_y,\vec{F}_z]$ a vector of fluxes and $\vec{S}$ a vector of sources. These elements for the system (\ref{eq: mass conservation})-(\ref{eq: momentum conservation}) read as follows

\begin{eqnarray}
\vec{u} &=& \left[ 
\begin{array}{c}
\rho\\
\rho v_x\\
\rho v_y\\
\rho v_z
\end{array}
\right],~~ 
\vec{F}_x = \left[ 
\begin{array}{c}
\rho v_x\\
\rho v_x v_x\\
\rho v_x v_y\\
\rho v_x v_z\\
\end{array}
\right],~~ 
\vec{F}_y = \left[ 
\begin{array}{c}
\rho v_y\\
\rho v_y v_x\\
\rho v_y v_y\\
\rho v_y v_z\\
\end{array}
\right]\nonumber\\
\vec{F}_z &=& \left[ 
\begin{array}{c}
\rho v_z\\
\rho v_z v_x\\
\rho v_z v_y\\
\rho v_z v_z\\
\end{array}
\right],~~
\vec{S} = \left[ 
\begin{array}{c}
0\\
-\rho \partial_x (V+Q)\\
-\rho \partial_y (V+Q)\\
-\rho \partial_z (V+Q)\\
\end{array}
\right]\label{eq:fluxbalance2}.
\end{eqnarray}

\noindent We calculate the numerical fluxes using the HLLE formula \cite{Thomas2}, according to which the fluxes at the left $\vec{u}_L$ and at the right $\vec{u}_R$ from each intercell boundary, along each of the three Cartesian directions are

\begin{equation}
\vec{F}_i = 
\dfrac{\lambda_{+}\vec{F}_i(\vec{u}_L) - \lambda_{-}\vec{F}_i(\vec{u}_R + \lambda_{+}\lambda_{-}(\vec{u}_R - \vec{u}_L))}{\lambda_{+}-\lambda_{-}},
\end{equation}

\noindent where $i=x,y,z$, $\lambda_{-}$ and $\lambda_{+}$ are the approximations of the characteristic velocities which are calculated as

\begin{eqnarray}
\lambda_- & = & \min(0, (v_i)_{L}, (v_i)_{R}), \\
\lambda_+ & = & \max(0, (v_i)_{L}, (v_i)_{R}).
\end{eqnarray}

Notice that with this approach we have not involved the pressure of the fluid, in this case the pressure-like tensor associated to the quantum fluid within the fluxes. Instead, the gradient of $Q$ is assumed to contribute as a source.

{\it Atmosphere.} An inherent ingredient of FV methods is the use of an atmosphere, which is defined as a minimum value of the fluid density, set to $\rho_{atm}$. In fluid dynamics it is useful to avoid the divergence of temperature $T=p/\rho$, with $p$ the pressure, which leads to the divergence of all other involved state variables. In our case there is no pressure, but there is the quantum potential $Q$ in (\ref{eq:qpotential}), which diverges or is undefined for $\rho=0$, unless $\nabla^2\sqrt{\rho}$ compensates the divergence.

{\it Boundary conditions.} In order to implement the isolation condition of the system, unlike the sponge in the SP frame,  we impose outflow boundary conditions on the fluid variables. In order to solve Poisson equation we use a monopolar boundary condition for the gravitational potential $V_{\partial D}=-M/r_{\partial D}$, where $r_{\partial D}$ is the distance from the origin to a given point of the numerical boundary and $M$ is given by Eq. (\ref{eq:mass}), and solve using the same Multigrid 2-level V-cycle method used for the SP frame.

\subsection{Evolution of an equilibrium configuration}

\begin{figure}
\centering
\includegraphics[width=7cm]{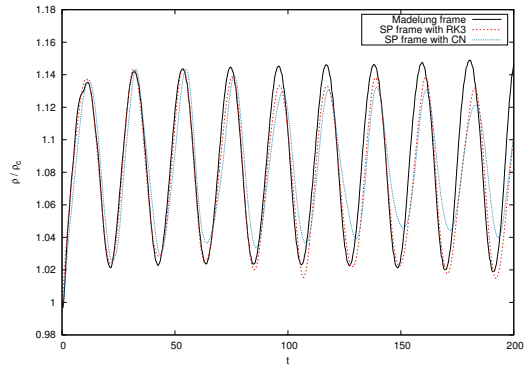} 
\caption{Central density $|\Psi(\vec{0},t)|^2$ for the solution in the SP frame and $\rho(\vec{0},t)$ in the Madelung frame, as functions of time. This plot shows that the first oscillation mode is similar, although not perfectly equal in the two frames. In the SP frame the period coincides with that in \cite{Guzman2019}, whereas the period in the Madelung frame shrinks in time.}
\label{fig:sph}
\end{figure}

In order to compare the essentials of ground state configurations we evolve the equilibrium configurations from Section \ref{sec:eqs} and track their behavior.  Using the numerical methods described above for the evolution, we integrate in time the equations for these configurations centered at the coordinate origin, and in Figure \ref{fig:sph} we show the central value of the density, $|\Psi(\vec{0},t)|^2$ in the SP frame and $\rho(\vec{0},t)$ in the Madelung frame. The result indicates that the configuration remains nearly stationary with an  oscillation mode consistent with the dominant spherical mode of the configuration with period $T=21.64$ as pointed out in \cite{Guzman2019} using the SP frame, with both the MoL and CN methods. When using the Madelung frame there is a reduction of the period in time.

\begin{figure}
\centering
\includegraphics[width=8cm]{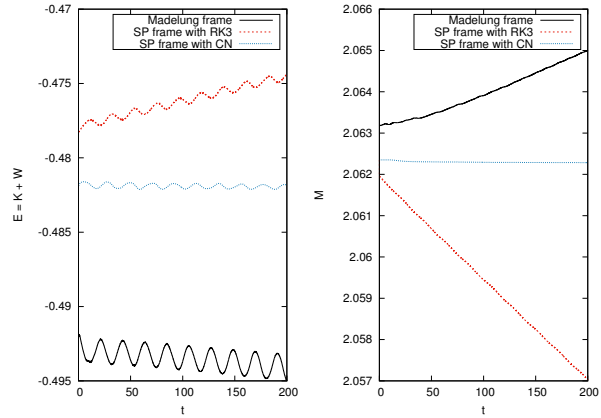} 
\caption{Evolution of the total energy $E=K+W$ and the mass $ M $ of the equilibrium configuration in the two frames.}
\label{fig:sph_data}
\end{figure}

In Figure \ref{fig:sph_data} we show the evolution of the mass $M$ and total energy $E=K+W$ as functions of time. 
We can see that in the Madelung frame the mass is slightly bigger than in the SP frame, which is due to the contribution of the atmosphere. Since this atmosphere is of  order of $ \rho_{atm} \sim 10^{-8} $, it contributes with a mass of order $M_{atm}\sim 10^{-3}$. On the other hand, the mass in the SP frame decreases when using the MoL, which is due to the dissipation of the time integration with the explicit RK3 method, whereas the conservation is better when using the CN method, indicating the evolution is closer to unitary.

Concerning the total energy, the values in the two frames disagree due to the difference in mass, which contributes to the gravitational energy $W$. For this reason in the Madelung frame the total energy $E$ is smaller than in the SP frame.

\subsection{Boosted equilibrium configuration}

A second comparison is dynamical and corresponds to a boosted equilibrium configuration. For this we apply an initial velocity $ v^0_x = 1 $ to the equilibrium configuration placed at the initial position $ (-10,0,0) $, in order to obtain a configuration traveling to the right along the $x$-axis. 

The linear momentum at initial time is applied to the equilibrium configuration in the SP frame as $\Psi(\vec{x},0) = \psi_{0}e^{{\rm i} v^0_x x}$, where $\psi_0$ is the wave function obtained from the solution of the eigenvalue problem (\ref{eq:SPSchro})-(\ref{eq:SPPoisson}). In the  Madelung frame the density is $\rho(\vec{x},0) = \rho_{0}$, and the initial speed is set to $\vec{v}(\vec{x},0 )=(v^0_x,0,0)$ where $\rho_0$ is that of the solution of the eigenvalue problem (\ref{eq: u prime})-(\ref{eq: u primeorigin}).

\begin{figure}[hbtp]
\centering
\includegraphics[width=7cm]{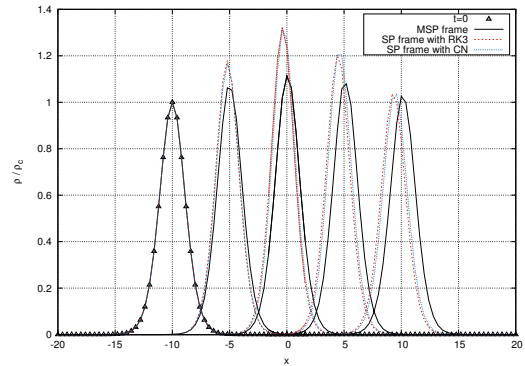}
\caption{Snapshots of the density profile every $ t = 5 $ time units in the Madelung frame (solid line) and the SP frame (dotted lines).}
\label{fig: rho-boost}
\end{figure}

Snapshots of the density are shown in Figure \ref{fig: rho-boost}. Notice that the maximum of the configuration using the hydrodynamical method in the Madelung frame locates at the appropriate position, since the velocity field is one. On the other hand, the configurations evolved in the SP frame retard with respect to the appropriate location of the maximum density. Nevertheless, this is due to the implementation of the momentum, which actually is a phase, and not a true velocity field at initial time.

\begin{figure}[hbtp]
\centering
\includegraphics[width=7cm]{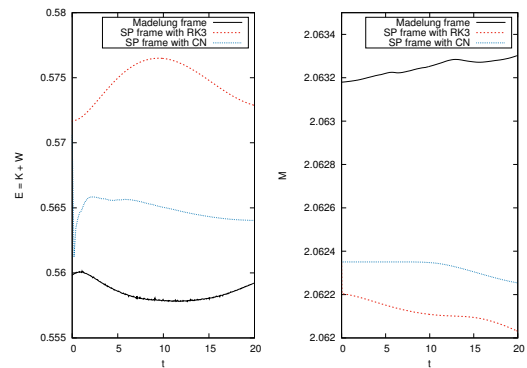}
\caption{Evolution of the total energy $E = K+W$and the mass $M$ as a function of time in the two frames for the boosted configuration.}
\label{fig: data-boost}
\end{figure}

In Figure \ref{fig: data-boost} we show the evolution of the total energy $ E = K + W $ and the mass $M$ in the two frames. Notice that the energy is well preserved in the two frames and the three methods. The mass decreases in a small fraction in the SP frame that we attribute to the sponge, whereas in the Madelung frame the mass remains relatively constant.

\section{Conclusions}
\label{sec:conclusions}

We have described the construction of ground state equilibrium configurations of the Schr\"odinger-Poisson system directly from the equations in the Madelung frame and verified that the solutions in both frames coincide within numerical accuracy.

Since the equations in the Madelung frame are similar to those of a fluid, we implement a simple scheme involving shock capturing methods used in fluid dynamics to simulate the evolution in time. We compare the evolution in SP and Madelung frames using the evolution equilibrium configuration when at rest and when the configuration is boosted.

We find that important limitations of the FV method arise. One of them is the need of an atmosphere used to avoid the singularity of the quantum potential at rarified regions. We notice that this also happens in simple cases within Quantum Mechanics, for example in excited states for a particle in a box or a harmonic oscillator potential, where the density has zeroes,  then using an atmosphere is needed to avoid the singularity $1/\sqrt{\rho}$ of $Q$. However the use of the atmosphere  introduces a non-derivable $\sqrt{\rho}$ that will spoil the factor $\nabla^2 \sqrt{\rho}$ of $Q$ in the intersection between the atmosphere and higher values of $\rho$. It seems that using solely the Madelung frame leads to a dilemma.

This type of problems seems to motivate the use of patches to the FV method, like in \cite{Hopkins_2019} where in low density regions the Smoothed Particle Hydrodynamics (SPH) method is used; or the use the SPH method to solve the evolution equations in the Madelung frame for $\rho$ and $\vec{v}$ in \cite{PhysRevE.91.053304}. It would be interesting to investigate how SPH develops in these simple scenarios.


\funding{IAR receives support within the CONACyT graduate scholarship program under the CVU 967478. This research is supported by grants CIC-UMSNH-4.9 and CONACyT Ciencias de Frontera Grant No. Sinergias/304001. The runs were carried out in the Big Mamma cluster of the Laboratorio de Inteligencia Artificial y Superc\'omputo, IFM-UMSNH.}

\dataavailability{The data underlying this article will be shared on reasonable request to the corresponding author.} 


\conflictsofinterest{The authors declare no conflict of interest. The funders had no role in the design of the study; in the collection, analyses, or interpretation of data; in the writing of the manuscript, or in the decision to publish the results'.} 



%

\begin{adjustwidth}{-\extralength}{0cm}

\reftitle{References}


\bibliography{BECDM}

\end{adjustwidth}
\end{document}